\def\be{\begin{equation}}
\def\ee{\end{equation}}

\def\Rb{{I\!\! R}}
\def\Ib{{I\!\! I}}

\def\cb{{\Bbb C}}

\def\ee{\end{equation}}
\def\ba{\begin{array}}
\def\ea{\end{array}}

\def\Ib{{I\!\! I}}

\def\Rb{{I\!\! R}}

\def\cb{{\Bbb C}}

\documentstyle[12pt]{article}
\input amssym.def
\begin{document}
\baselineskip18pt \thispagestyle{empty}
\begin{center}
{\Large \bf  On Integrability and Pseudo-Hermitian Systems}

{\Large \bf  with Spin-Coupling Point Interactions}
\end{center}
\vskip 2mm

\begin{center}
{\normalsize Shao-Ming Fei}
\end{center}

\begin{center}
\begin{minipage}{5.3in}
{\small \sl Department of Mathematics, Capital  Normal
University, Beijing, China}

{\small \sl Institute of Applied Mathematics, University of Bonn, D-53115 Bonn, Germany}

\end{minipage}
\end{center}

\begin{center}
\begin{minipage}{5.2in}
\vskip 3mm {\bf Abstract}

We study the pseudo-Hermitian systems with general spin-coupling
point interactions and give a systematic description of the
corresponding boundary conditions for PT-symmetric systems.
The corresponding integrability for both
bosonic and fermionic many-body systems with PT-symmetric contact interactions
is investigated.

\vskip 1mm
PACS number(s): 02.30.Ik, 11.30.Er, 03.65.Fd
\end{minipage}
\end{center}

\bigskip
\bigskip

Self-adjoint quantum mechanical models describing a particle moving in a
local singular potential have been extensively discussed
\cite{agh-kh1,agh-kh2,agh-kh3}. The integrability of one dimensional many-body systems
with self-adjoint contact interactions has been studied according to
Yang-Baxter relations \cite{adf}. The results are generalized
to the case of particles with spin-coupling interactions
\cite{afk}.

PT-symmetric quantum mechanical models have been studied from some
mathematical and physical considerations \cite{pt}.
In \cite{pt-fei} the classification and spectra problem of
PT-symmetric point interactions are investigated.
The integrability of many-body systems with PT-symmetric
interactions is clarified \cite{pt-int}. The $\delta$-type
spin-coupling interactions with PT-symmetry is discussed in
\cite{pt-spin}.

In this letter we study the boundary conditions for PT-symmetric
point interactions of particles with spin-coupling, and the integrability of
bosonic and fermionic many-body systems with PT-symmetric,
spin-coupling contact interactions characterized by these boundary
conditions.

One dimensional quantum mechanical models of spinless particles with
point interactions at the origin can be characterized by separated
or nonseparated boundary conditions imposed on the (scalar) wave
function $\varphi$ at $x=0$. The family of point interactions for
the Schr\"odinger operator $ - \frac{d^2}{dx^2}$ can be described by
unitary $ 2 \times  2 $ matrices via von Neumann formulas for
self-adjoint extensions of symmetric operators:
\begin{equation} \label{bound}
\left( \begin{array}{c}
\varphi\\
\varphi '\end{array} \right)_{0^+}
= e^{i\theta} \left(
\begin{array}{cc}
a & b \\
c & d \end{array} \right)
\left( \begin{array}{c}
\varphi\\
\varphi '\end{array} \right)_{0^-},
\end{equation}
where $ad-bc = 1$, $\theta, a,b,c,d \in \Rb$.
$\varphi(x)$ is the scalar wave function of two spinless particles with
relative coordinate $x$. (\ref{bound}) also describes two particles
with spin $s$ but without any spin coupling between the particles when they
meet, in this case $\varphi$ represents any one of the
components of the wave function.

The separated boundary conditions with respect to
self-adjoint interactions are described by
\be\label{bounds2}
\varphi^\prime(0_+) = h^+ \varphi (0_+)~, ~~~
\varphi^\prime(0_-) = h^- \varphi (0_-),
\ee
where $h^{\pm} \in \Rb \cup \{ \infty\}$.
$ h^+ = \infty$ or $ h^- = \infty$
correspond to Dirichlet boundary conditions and
$ h^+ = 0$ ~or~ $ h^- = 0$ correspond to
Neumann boundary conditions.

The family of PT-symmetric point interactions
is described by the boundary conditions at the origin of one of the
following two types
\begin{equation} \label{bcond1} \left(\begin{array}{c}
\varphi \\
\varphi'
\end{array}\right)_{0_+} = e^{i \theta}
\left( \begin{array}{cc}
\sqrt{1 +bc} \; e^{i\phi} & b \\
c & \sqrt{1+bc} \; e^{-i\phi}
\end{array} \right)
 \left(\begin{array}{c}
\varphi  \\
\varphi'
\end{array}\right)_{0_-};
\end{equation}
with the real parameters $ b \geq 0, c \geq -1/b$ (if the
parameter $b$ is equal to zero, then the last inequality can be
neglected), $ \theta, \phi \in [0, 2 \pi);$ and
\begin{equation} \label{bcond2}
h_0 \varphi' (0_+)  =  h_1 e^{i\theta} \varphi (0_+)\,,~~~
h_0 \varphi' (0_-)  =  - h_1 e^{-i\theta} \varphi (0_-)
\end{equation}
with the real phase parameter $ \theta \in [0,2\pi)$ and with the
parameter $ {\bf h} = (h_0, h_1)$ taken from the (real) projective
space $ {\bf P}^1.$

For a particle with spin $s$, the wave function has $n=2s+1$ components.
Therefore two particles with contact interactions have a general boundary
condition described in the center of mass coordinate system by:
\be\label{BOUND}
\left( \begin{array}{c}
\psi\\
\psi '\end{array} \right)_{0^+}
=\left(
\begin{array}{cc}
A & B \\
C & D \end{array} \right)
\left( \begin{array}{c}
\psi\\
\psi '\end{array} \right)_{0^-},
\end{equation}
where $\psi$ and $\psi '$ are $n^2$-dimensional
column vectors, $A,B,C$ and $D$ are
$n^2\times n^2$ matrices. The boundary condition (\ref{BOUND})
includes not only the usual contact interaction between the particles,
but also the spin couplings of the two particles.

For self-adjoint point interactions, due to
the required symmetry condition of the Schr\"odinger operator:
$$
\displaystyle<-\frac{d^2}{dx^2}u,v>_{L_2(\Rb,\cb^n)}=
<u,-\frac{d^2}{dx^2}v>_{L_2(\Rb,\cb^n)},
$$
for any $u,v\in C^\infty(\Rb \setminus \{0\})$,
the matrices $A,B,C$, and $D$ are subject to
the following restrictions:
\be\label{ABCD}
A^\dagger D-C^\dagger B=\Ib,~~~B^\dagger D=D^\dagger B,~~~
A^\dagger C=C^\dagger A,
\ee
where $\dagger$ stands for the conjugate and transpose.

Corresponding to (\ref{bounds2}) the separated boundary conditions are given by
\be\label{BOUNDS}
\psi^\prime(0_+) = G^+ \psi (0_+), ~~~
\psi^\prime(0_-) = G^- \psi (0_-) ,
\ee
where, for the self-adjoint point interactions, $G^{\pm}$ are Hermitian matrices.

We consider now the boundary conditions describing the point interactions
with PT-symmetric and spin-coupling interactions.
By applying PT-operation to (\ref{BOUND}), we have
$$
\left( \begin{array}{c}
\psi\\
\psi '\end{array} \right)_{0^-}
=\left(
\begin{array}{cc}
A^\ast & -B^\ast \\
-C^\ast & D^\ast \end{array} \right)
\left( \begin{array}{c}
\psi\\
\psi '\end{array} \right)_{0^+},
$$
where $\ast$ stands for conjugation.
Hence $A$, $B$, $C$, $D$ satisfy
\be\label{ABCD1}
AA^\ast-BC^\ast=\Ib,~~
DD^\ast-CB^\ast=\Ib,~~
BD^\ast=AB^\ast,~~
CA^\ast=DC^\ast,
\ee
where $\Ib$ is the $n^2\times n^2$ identity matrix.
The boundary condition (\ref{BOUND}) with $A$, $B$, $C$, $D$ satisfying
(\ref{ABCD1}) represents PT-symmetric point interactions with
spin coupling.

Accordingly the separated type boundary conditions
with respect to PT-symmetric interactions are given by
\be\label{boundsep}
\psi^\prime(0_+) = F \psi (0_+), ~~~
\psi^\prime(0_-) = G \psi (0_-) ,
\ee
where $G=-F^\ast$.

The case $A=D=\Ib$, $B=0$, $C=C^\ast$
corresponds to a Hamiltonian with PT-symmetric $\delta$-type
interactions (when $C$ is further symmetric, the system is both
PT-symmetric and self-adjoint). The case $A=D=\Ib$, $C=0$, $B=B^\ast$
corresponds to a PT-symmetric Hamiltonian $H$ of the form:
$$
H=-D^2_x(1+B\delta) - BD_x\delta^\prime,
$$
where $D_x$ is defined by $(D_x f)(\varphi)=-f(\frac{d}{dx}\varphi)$, for
$f\in C^\infty_0(\Rb\slash\{0\})$ and $\varphi$ a test function with a
possible discontinuity at the origin. $H$ is
self-adjoint when $B$ is Hermitian \cite{afk}, and $H$ is both self-adjoint
and PT-symmetric when $B$ is real symmetric.

Concerning the integrability of many-body systems with spin-coupling interactions,
we first consider two-particle case.
Let $e_\alpha$, $\alpha=1,...,n$,
be the basis (column) vector with the $\alpha$-th component as $1$
and the rest components $0$. The wave function of the system
is of the form
\be\label{psi}
\psi=\sum_{\alpha,\beta=1}^n\phi_{\alpha
\beta}(x_1,x_2)e_\alpha\otimes e_\beta.
\ee

According to the statistics
$\psi$ is symmetric (resp. antisymmetric) under the interchange of
the two particles if $s$ is an integer (resp. half integer).
Let $k_1$ and $k_2$ be the momenta of the two particles.
In the region $x_1<x_2$, in terms of Bethe hypothesis the
wave function has the following form
\be\label{w1}
\psi=u_{12}e^{i(k_1x_1+k_2x_2)}+u_{21}e^{i(k_2x_1+k_1x_2)}
=u_{12}e^{i(K_{12} X-k_{12}x)}+u_{21}e^{i(K_{12}X+k_{12}x)},
\ee
where $X=(x_1+x_2)/2$, $x=x_2-x_1$ are the coordinates of the
center of mass system, $K_{12}=k_1+k_2$, $k_{12}=(k_1-k_2)/2$,
$u_{12}$ and $u_{21}$ are $n^2\times 1$ column matrices.

In the region $x_1>x_2$,
\be\label{w2}
\psi=(P^{12}u_{12})e^{i(K_{12} X+k_{12}x)}
+(P^{12}u_{21})e^{i(K_{12} X-k_{12}x)},
\ee
where according to the symmetry or antisymmetry conditions,
$P^{12}=p^{12}$ for bosons and $P^{12}=-p^{12}$ for fermions, $p^{12}$
being the operator on the $n^2\times 1$ column that interchanges
the spins of the two particles.
Substituting (\ref{w1}) and (\ref{w2}) into the boundary
conditions (\ref{BOUND}), we get
\be\label{a1}
\left\{
\begin{array}{l}
u_{12}+u_{21}
=A\,P^{12}(u_{12}+u_{21})+ik_{12}B\,P^{12}(u_{12}-u_{21}),\\[3mm]
ik_{12}(u_{21}-u_{12})
=C\,P^{12}(u_{12}+u_{21})+ik_{12}D\,P^{12}(u_{12}-u_{21}).
\end{array}\right.
\ee
Eliminating the term $P^{12}u_{21}$ from
(\ref{a1}) we obtain the relation
\be\label{2112}
u_{21} = Y_{21}^{12} u_{12}~,
\ee
where
\be\label{a21a12}
\ba{rcl}
Y_{21}^{12}&=&[(A-ik_{12}B)^{-1}-ik_{12}(C-ik_{12}D)^{-1}]^{-1}\\[3mm]
&&[(A-ik_{12}B)^{-1}(A+ik_{12}B)P^{12}
-(C-ik_{12}D)^{-1}(C+ik_{12}D)P^{12}\\[3mm]
&&-(A-ik_{12}B)^{-1}-ik_{12}(C-ik_{12}D)^{-1}].
\ea
\ee

Similarly, with respect to the separated type boundary
condition (\ref{boundsep}), we have
\be\label{a1sep}
\ba{l}
ik_{12}(u_{21}-u_{12})=F(u_{12}+u_{21}),\\[3mm]
ik_{12}P^{12}(u_{12}-u_{21})=-F^\ast\,P^{12}(u_{12}+u_{21}).
\ea
\ee
The two equations above are compatible when $F$ satisfies
\be\label{compa}
(ik_{12}-F)^{-1}(ik_{12}+F)=P^{12}(ik_{12}-F^\ast)^{-1}(ik_{12}+F^\ast)P^{12}.
\ee
(\ref{compa}) is satisfied when $F$ commutes with $P^{12}$, which implies
that $F$ is real. For the case of spin-$1\over 2$ ($n=2$), $F$ is generally of the form
\be\label{hspin}
F=\left(\ba{cccc}a&e_1&e_1&c\\
e_3&f&g&e_2\\e_3&g&f&e_2\\
d&e_4&e_4&b\ea\right),
\ee
where $a,b,c,d,f,g,e_1,e_2,e_3,e_4\in\Rb$.
In stead of (\ref{a21a12}), from (\ref{a1sep}) we have, upon to the
condition (\ref{compa}),
\be\label{a21a12sep}
Y_{21}^{12}=(ik_{12}-F)^{-1}(ik_{12}+F).
\ee

In the following we consider the integrability of systems with
point interaction described by the boundary condition (\ref{boundsep}).
For a system of $N$ identical particles with PT-symmetric contact interactions
characterized by the separated type operator (\ref{a21a12sep}),
the wave function in a given region, say $x_1<x_2<...<x_N$, is of the form
\be\label{Psi}
\ba{rcl}
\Psi&=&\displaystyle\sum_{\alpha_1,...,\alpha_N=1}^n
\phi_{\alpha_1,...,\alpha_N}(x_1,...,x_N)
e_{\alpha_1}\otimes...\otimes e_{\alpha_N}\\[5mm]
&=&u_{12...N}e^{i(k_1x_1+k_2x_2+...+k_Nx_N)}
+u_{21...N}e^{i(k_2x_1+k_1x_2+...+k_Nx_N)}\\[3mm]
&&+(N!-2)\, {\rm other~terms},
\ea
\ee
where $k_j$, $j=1,...,N$, are the momentum of the $j$-th particle.
$u$ are $n^N\times 1$ matrices.
The wave functions in the other regions are determined
from (\ref{Psi}) by the requirement of
symmetry (for bosons) or antisymmetry (for fermions).
Along any plane $x_i=x_{i+1}$, $i\in 1,2,...,N-1$, we have
\be\label{a1n}
u_{\alpha_1\alpha_2...\alpha_j\alpha_{j+1}...\alpha_N}=
Y_{\alpha_{j+1}\alpha_j}^{jj+1}
u_{\alpha_1\alpha_2...\alpha_{j+1}\alpha_j...\alpha_N},
\ee
where
\be\label{y}
Y_{\alpha_{j+1}\alpha_j}^{jj+1}=[ik_{\alpha_j\alpha_{j+1}}-F_{j{j+1}}]^{-1}
[ik_{\alpha_j\alpha_{j+1}} + F_{jj+1}].
\ee
Here $k_{\alpha_j\alpha_{j+1}}=(k_{\alpha_j}-k_{\alpha_{j+1}})/2$ play
the role of momenta
and $P^{jj+1}=p^{jj+1}$ for bosons and $P^{jj+1}=-p^{jj+1}$ for fermions,
where $p^{jj+1}$ is the operator on the $n^N\times 1$ column $u$
that interchanges the spins of particles $j$ and $j+1$. $F_{jj+1}$
stands for the application of the operator $F$ to the $j$th and $j+1$th particles.

For consistency $Y$ must satisfy the Yang-Baxter equation with
spectral parameter \cite{y},
\be\label{ybe1}
Y^{m,m+1}_{ij}Y^{m+1,m+2}_{kj}Y^{m,m+1}_{ki}
=Y^{m+1,m+2}_{ki}Y^{m,m+1}_{kj}Y^{m+1,m+2}_{ij}.
\ee
It is straight forward to verify that the operator $Y$ given by (\ref{y}) satisfies
the Yang-Baxter equation (\ref{ybe1}).
Therefore this system is integrable with the exact wave functions given by
(\ref{Psi}).

As $F$ is a matrix satisfying (\ref{compa}), it's spectra are not
real in general. For instance, the real matrix (\ref{hspin})
gives complex eigenvalues. Hence the PT-symmetric (separated type) contact
couplings contain non-Hermitian interactions with complex and real spectra,
while the real spectra, e.g. spectra of real $F$ commuting with $P^{12}$, covers
part of the real spectra from the Hermitian interactions, when the matrix
$F$ is further symmetric ($e_3=e_1$, $e_4=e_2$, $d=c$ in the case of (\ref{hspin})).

As for bound states, let us assume that $F$ has real spectra.
Let $\Gamma$ be the set of $n^2$ eigenvalues of $F$. For any
$\lambda_\alpha\in \Gamma$ such that $\lambda_\alpha<0$, there are
$2^{N(N-1)/2}$ bound states for the $N$-particle system,
\be\label{bpsins}
\psi^{N}_{\alpha\underline{\epsilon}}=
v_{\alpha\underline{\epsilon}}
\prod_{k>l} (\theta (x_k-x_l) +\epsilon_{kl}\theta (x_l-x_k))
e^{\lambda_\alpha\sum_{i>j} \vert x_i-x_j\vert },
\ee
where $v_{\alpha\underline{\epsilon}}$ is the spin wave function and
$\underline{\epsilon} \equiv \{ \epsilon_{kl}\,:\,k>l \}$; $\epsilon_{kl}=\pm$,
labels the $2^{N(N-1)/2}$-fold degeneracy.
The spin wave
function $v$ here satisfies $P^{ij}v_{\alpha\underline{\epsilon}}
=\epsilon_{ij}v_{\alpha\underline{\epsilon}}$
for any $i\neq j$, that is, $p^{ij}v_{\alpha\underline{\epsilon}}
=\epsilon_{ij}v_{\alpha\underline{\epsilon}}$
for bosons and $p^{ij}v_{\alpha\underline{\epsilon}}
=-\epsilon_{ij}v_{\alpha\underline{\epsilon}}$ for fermions.
The energy of the bound state $\psi^{N}_{\alpha\underline{\epsilon}}$ is
\be\label{es}
E_\alpha=-\frac{\lambda_\alpha^2}{3}N(N^2-1).
\ee

We have studied the boundary conditions for systems
with general PT-symmetric spin-coupling point interactions,
and the corresponding integrability for both
bosonic and fermionic many-body systems with separated-type
PT-symmetric contact interactions (\ref{a21a12sep}).
The scattering matrices can be also studied similar
to the case of self-adjoint interaction case.
The spectra and integrability for many-body systems with
PT-symmetric contact interactions described by the $Y$
operator (\ref{a21a12}) for the non-separated boundary
conditions can be studied accordingly
in terms of the Yang-Baxter equation (\ref{ybe1}), though
it could be quite complicated as $A$, $B$, $C$, $D$ subject
to the conditions (\ref{ABCD1}).


\begin{thebibliography}{99}

\bibitem{agh-kh1}
S. Albeverio, F. Gesztesy, R. H\o egh-Krohn and H. Holden, {\it
Solvable Models in Quantum Mechanics}, New York: Springer, 1988.

\bibitem{agh-kh2}
M. Gaudin, {\it La fonction d'onde de Bethe}, Masson, 1983.

\bibitem{agh-kh3}
S.Albeverio and P.Kurasov,
{\it  Singular perturbations of differential operators. Solvable
Schr\"odinger type operators},
 London Mathematical Society Lecture Note Series, 271,
  Cambridge University Press, Cambridge, 2000.

\bibitem{adf}
S. Albeverio, L. D{\c a}browski and S.M. Fei,
Int. J. Mod. Phys. B {\bf 14}, 721-727 (2000).

\bibitem{afk}
S. Albeverio, S.M. Fei and P. Kurasov, Rep. Math. Phys. {\bf 47},
157-165 (2001).

\bibitem{pt}
C.M.Bender and S.Boettcher, {\it Phys. Rev. Lett.},{\bf 80} (1998),
5243--5246.\\
C.M.Bender, S.Boettcher, H.F.Jones, and V.M.Savage,
{\it J. Phys. A}, {\bf 32} (1999), 6771--6781.\\
C.M.Bender, G.V.Dunne, and P.N.Meisinger, {\it
Phys. Lett. A},{\bf 252} (1999), 272--276.\\
G.L\'evai and M.Znojil, {\it J. Phys. A}, {\bf 33}
(2000), 7165--7180.\\
M.Znojil, {\it J. Phys. A}, {\bf 33} (2000),  4561--4572.\\
M.Znojil, F.Cannata, B.Bagchi, and R.Roychoudhury,
{\it Phys. Lett. B}, {\bf 483} (2000),  284--289.\\
M.Znojil and M.Tater, {\it J. Phys. A}, {\bf 34} (2001), 1793--1803.\\
A.Mostafazadeh, {\it Class. Quantum Grav.} {\bf 20}(2003)155.\\
Andrianov A A, Cannata F, Dedonder J P and Ioe M V, {\it Int. J. Mod.
Phys.} {\bf A 14}(1999)2675.\\
Klishevich S M and Plyushchay M S, {\it Nucl. Phys.} {\bf B 616}(2001)403.\\
Mostafazadeh, {\it Nucl. Phys.} {\bf B 640}(2002)419.\\
Znojil M, {\it Nucl. Phys.} {\bf B 662}(2003)554.\\
Hatano N and Nelson D R, {\it Phys. Rev. Lett.} {\bf 77}(1996)570.\\
Mostafazadeh A, {\it Czech. J. Phys.} {\bf 54}(2004)93.\\
Bender C M, Brody D C and Jones H F, {\it Phys. Rev. Lett.} {\bf 89}(2002)270401.

\bibitem{pt-fei}
S. Albeverio, S.M. Fei and P. Kurasov,
Lett. Math. Phys {\bf 59}, 227-242 (2002).

\bibitem{pt-int}
S.M. Fei, {\sf Czech J. Phys.} {\bf 54}(2004)43-49.

\bibitem{pt-spin}
S.M. Fei, {\sf Czech J. Phys.} {\bf 53}(2003)1027-1034.

\bibitem{y}
C.N. Yang, Phys. Rev. Lett. {\bf 19}(1967)1312-1315.\\
C.N. Yang, Phys. Rev. {\bf 168}(1968)1920-1923.\\
C.H. Gu and C.N. Yang, Commun. Math. Phys. {\bf 122}
(1989)105-116.
\end{thebibliography}
\end{document}